\documentclass[prd]{revtex4}
\usepackage{graphicx, epsfig}
\usepackage{color}
\usepackage{mathrsfs}

\newcommand{\be}{\begin{equation}}
\newcommand{\ee}{\end{equation}}
\newcommand{\bea}{\begin{eqnarray}}
\newcommand{\eea}{\end{eqnarray}}

\newcommand{\gapp}{\mathrel{\raise.3ex\hbox{$>$}\mkern-14mu
              \lower0.6ex\hbox{$\sim$}}}
\newcommand{\lapp}{\mathrel{\raise.3ex\hbox{$<$}\mkern-14mu
              \lower0.6ex\hbox{$\sim$}}}

\begin{document}
\title{Using Cosmology to Establish the Quantization of Gravity}
\author{Lawrence M. Krauss}
\affiliation{School of Earth and Space Exploration and Department of Physics, Arizona State University, Tempe AZ 85287-1404}
\affiliation{and, Mount Stromlo Observatory, Research School of Astronomy and Astrophysics, Australian National University, Weston, ACT, Australia, 2611}
\author{Frank Wilczek}
\affiliation{Center for Theoretical Physics, MIT, Cambridge MA 02139}

\begin{abstract}
\noindent
{\bf
While many aspects of general relativity have been tested, and general principles of quantum dynamics demand its quantization, there is no direct evidence for that.   It has been argued that development of detectors sensitive to individual gravitons is unlikely, and perhaps impossible.  We argue here, however, that measurement of polarization of the Cosmic Microwave Background due to a long wavelength stochastic background of gravitational waves from Inflation in the Early Universe would firmly establish the quantization of gravity.}

\end{abstract}

\maketitle

Direct detection of gravitational waves is an exciting frontier of experimental physics, with positive results anticipated soon (i.e. \cite{Aasi:2013nha}).  The anticipated signals are classical disturbances, comprised of coherent superpositions of many individual quanta.   The possibility of detecting individual gravitons is far more daunting.  Indeed, recently Freeman Dyson and colleagues \cite{Rothman:2006} have cogently estimated that it may in fact be infinitely more daunting, namely that  it is likely to be impossible to physically realize a detector sensitive to individual gravitons without having the detector collapse into a black hole in the process.  

If that is the case, one might wonder whether we can ever directly validate any quantum effects associated with the gravitational field.   That would be ironic, not to say pathetic, since the apparent tension between quantum mechanics and a full quantum treatment of general relativity has been one of the driving forces in much of fundamental particle theory over the past 30 years.  

The purpose of this note is to point out that cosmology provides a realistic observable that is directly tied to the quantization of gravity.  Specifically, observation of a cosmological gravitational wave background associated with an inflationary phase would provide, as a bonus, compelling evidence for the quantization of the gravitational field.  It does so in a way which is at least heuristically equivalent to all laboratory experiments that probe quantum phenomena--it couples quantum mechanical phenomena to a classical detector, effectively amplifying quantum mechanical effects so that they are classically measurable.  The classical detector, in this case, is the expanding Universe.

Let us emphasize at the start that such a cosmological background has not yet been observed, and that its predicted magnitude, even within the inflationary scenario, depends on the rate of expansion during inflation.  If the background is not observed, it could simply indicate a relatively small rate of expansion.  But detection is a plausible possibility, as we describe, and major efforts are underway to achieve it.   We should also emphasize that no essentially new predictions or calculations are presented here; we are merely bringing to the fore an implication of existing results that seems particularly noteworthy.  

The fact that quantization associated with gravity appears to be an essential feature of a gravitational wave background generated by inflation is suggested by existing calculations, including the following.   A period of inflation in the early universe results from a period of quasi-de Sitter expansion associated with a scalar field in an almost flat potential.  If one considers a quantized approximately massless scalar field in de-Sitter space, expanded into Fourier components with quantized mode functions, $v_k$ then it is straightforward to calculate the zero-point quantum fluctuations of these mode functions,

\begin{equation}
\left< v_{\bf k}v_{\bf k'}\right> = P_v(k) \delta(\bf{k+k'})
\end{equation}
where, on large scales the Power Spectrum $P_v(k)$ approaches 
\begin{equation}
P_v = {1\over {2k^3}} (aH)^2
\end{equation}
where $a$ is the scale factor during the de-Sitter expansion, and $H$ is the Hubble expansion parameter associated with the de Sitter phase.

Now consider the two helicity states of transverse traceless metric perturbations, which we traditionally associate with classical gravitational waves.  As first pointed out by Grishchuk in 1975 \cite{Grishchuk:1974ny}, the Fourier modes of these two states, $h^{\rm{x},+}_k$ are each governed by an action in de Sitter space that is identical to that of a massless scalar,  with the  correspondence 
\begin{equation}
h_k={2 \over {aM_{pl}}} v_k
\end{equation}
Thus {\it if\/} one treats these Fourier modes as quantum modes, then there will be zero-point fluctuations in each of the two modes that can be directly derived from equation (2), leading to a power spectrum 
\begin{equation}
P_t = {4\over {k^3}} {H^2 \over M_{pl}^2}
\end{equation}
Once these modes leave the horizon during the Inflationary expansion, they freeze in, effectively amplifying the mode number while outside the horizon, and they return inside the horizon as a coherent superposition of many quanta, i.e. as a classical wave.   These waves, originating as quantum fluctuations, then have a dimensionless power spectrum at the horizon , given by 
\begin{equation}
\Delta^2(k) = {k^3 \over {2 \pi^2}} P_t = {2\over {\pi^2}} {{H^2 \over M_{pl}^2} }
\end{equation}
In this calculation the initial mode number is small, thus implicating quantum gravity.   

While the fact that this calculation relies on mode occupation originating in quantum fluctuations suggests that the calculated effect is essentially quantum-mechanical, that conclusion is not logically forced.   After all, many -- in principle, all! --  classical effects can also be calculated quantum mechanically, and sometimes that approach is even more direct or simpler.  Our claim that a gravitational wave background from inflation requires quantum effects in gravity for its generation can, however, be based on more general and perhaps firmer ground, without recourse to the specific calculation outlined above, using simple dimensional analysis.
 
In the de Sitter limit, the inflationary epoch is characterized by a single parameter, the Hubble parameter $H$.  Abstracting $M, L, T$ as dimensions of mass, length, and time, we therefore have
\begin{equation}
\left[ H \right] ~=~ \frac{1}{T}
\end{equation}
{\bf (A bracketed quantity represents the dimensional content of that quantity.)}
A contemporary gravitational wave background that was produced during the inflationary epoch will require gravitational interactions, and thus will involve the gravitational constant $G$.   {\bf We assume that the background density can be usefully expanded as an analytic function of the coupling, as it would appear in any perturbative approach to quantization.   We also note that the dimensionless ratio $G \hbar H^2 / c^5$ is small for sub-Planckian inflation, i.e. inflation with curvature scale less than the Planck length, while super-Planckian inflation is theoretically dubious.}  The lowest-order effect, which ({\bf if non-zero\/}) will dominate, therefore involves %just 
one power of $G$.   Now if we want to form a dimensionless numerical measure of the strength of the gravitational background, we should take into account the following circumstance.  The energy density {\bf $\rho_{\rm grav.} $  in gravitational radiation after inflation ends} gives a physical measure of the strength of the background, but it varies {\bf afterwards} with the length-scale $a$ of the expanding universe as $1/a^4$.  If we want to extract a relic of the early universe, we must compensate that factor.  So we will look to combine $G$ to the first power, together with powers of $H$ and the fundamental constants $\hbar, c$, and $L^4$, to produce an dimensionless invariant measure of the magnitude of the background.  Thus we require
\begin{equation}
\left[ G \right] \left[ H \right]^\alpha \left[ \hbar \right]^\beta \left[ c \right]^\gamma ~=~ {\bf  [ \rho_{\rm grav.} ] [ a^4 ]  ~=~ } \frac{\left[ E \right]}{L^3} \, L^4 ~=~ \frac{ML^3}{T^2}
\end{equation}
This has a unique solution $\alpha = 2, \beta = 2, \gamma = -4$.  Note that if factors of $\hbar$ and $c$ are made explicit in Eq. (5) then our dimensional analysis is vindicated.

Thus the gravitational radiation background, measured invariantly, is proportional to $\hbar^2$.   Since this is a positive power of $\hbar$, we infer the essentially quantum-mechanical nature of that phenomenon.  Since no field other than gravity is involved, we infer that quantization of the gravitational field is an essential ingredient.  It is instructive to compare this result for graviton radiation in cosmology with results for photon radiation in atomic physics.  $\hbar$ typically appears with a {\it negative \/} power in the decay rate of low-lying atomic levels.   The point is, that those levels themselves cannot be specified classically.  Radiation from classical ``Rydberg'' orbits is classical, and contains no powers of $\hbar$; however there is no classical gravitational radiation from a classical de Sitter background, and what radiation there is brings in positive powers of $\hbar$.

Inflation also in general predicts an almost flat spectrum of gaussian adiabatic primordial density fluctuations at the horizon, due to quantum fluctuations in the scalar field driving inflation, which can generate all observed structure in the Universe, and which appears to be in excellent quantitative agreement with observations of primordial temperature perturbations in the Cosmic Microwave Background (CMB).
If the inflation scale, $H$, is sufficiently large, horizon-sized gravitational waves will also produce measurable CMB effects \cite{Krauss:1992ke,Kamionkowski:1997av,Baumann:2008aq,Krauss:2010ty}.    For inflation with a single scalar field, the ratio of the polarization power due to these gravitational wave perturbations to the power associated with temperature (i.e. scalar density) fluctuations,  then (i.e. \cite{Baumann:2008aq}):
\begin{equation}
r = 0.01 {H_{inf}^2 \over(2.5 \times 10^{13} \, {\rm GeV})^2}.
\end{equation}
Observations currently give an upper limit on this ratio to be $ r < 0.11$ \cite{Komatsu:2010fb}, and it is possible that observations will be able to probe values of $r$ that are far smaller (i.e. \cite{Book:2011dz}).  
Thus, a gravitational wave background due to inflation associated with the scale suggested by coupling constant unification \cite{Dimopoulos:1981yj,Krauss:1992ke}, which corresponds to $H\approx 2.5 \times 10^{13}$ GeV, could be observed in the near future.  
While the current observations of CMB temperature fluctuations, and the observed flatness of the universe are strongly suggestive of an inflationary origin, the mere observation of polarization in the CMB compatible with a gravitational wave background, as exciting as that may be, will not alone prove that it originates in quantum phenomena associated with gravitation (i.e. \cite{Krauss:1991qu,JonesSmith:2007ne}).  Fortunately, there are a wide variety of consistency tests that can be performed to check for an inflationary origin (i.e. see \cite{Liddle:1992wi}).  These include a simple relationship between this ratio and the slope of the CMB temperature fluctuation power spectrum as a function of frequency.  In addition, inflation predicts super-horizon size correlations in the gravitational wave spectrum that might be discernible (i.e. see \cite{Baumann:2009mq}).  

If these consistency tests were satisfied quantitatively, we would thereby have reasonably unambiguous evidence that inflation did indeed occur, and that that linearized fluctuations in the gravitational field are quantized, with the power spectrum originating in quantum zero-point fluctuations in the gravitational field.  

We should contrast the joint appearance of $G$ and $\hbar$ in Eqs. (7) and (8), which really does implicate quantization of the gravitational field, with other cases, including specifically neutron interferometry, in which both appear.  These have sometimes been put forward as ``quantum gravitational'' phenomena, but more properly they are manifestations of the ordinary quantum mechanics of particles (i.e., neutrons) in {\it classical\/} gravitational fields.   Indeed it is more natural to express the effect in terms of the quantity $g$, the gravitational acceleration near Earth's surface, which is the relevant aspect of the experimental environment, and then $G$, which indicates intrinsically gravitational dynamics, does not appear at all.    Similar remarks apply to scalar mode perturbations within inflationary models.   

It is also possible, of course, that a fully realized theory of quantum gravity would have other indirect consequences that could be observed, {\it e. g.\/} the existence of unusual interactions, or even that it would dictate the entirety of a ``theory of everything''.  Perhaps the most concrete ideas along these lines arise in  gravity-mediated supersymmetry breaking, wherein quantum gravity effects make dominant contributions to the masses of supersymmetric particles \cite{Hall:1983iz,Soni:1983rm,Kawamura:1994ys}.  But those possibilities remain highly speculative.

Through inflation, the Universe can act effectively as a graviton detector built on an ``impractical scale''.  It amplifies a quantum mechanical effect to where it can be detected as a classical, observable signal, and may provide compelling empirical support for the quantization of gravity.   Thus we both illustrate and transcend, rather than contradict, the arguments of \cite{Rothman:2006}.

\acknowledgments

We are grateful to Freeman Dyson for stimulating our interest in this question and to the organizers of the 90th birthday celebration for Dyson at the NTU in Singapore, where he lectured on this subject.   We also thank Andrew Long, Subir Sabharwal, Tanmay Vachaspati for useful discussions and Freeman Dyson, Steve Weinberg, Edward Witten and Xerxes Tata for comments on early drafts of this manuscript. LMK is supported by the U.S. Department of Energy at ASU, and also by Australian National University.  FW supported by the U.S. Department of Energy under contract No. DE-FG02-05ER41360.

\bibliography{gravitons}
\end{document}